\documentclass[12pt]{article}%
\usepackage{amsmath}
\usepackage{amsfonts}
\usepackage{amssymb}
\usepackage{graphicx}%
\providecommand{\U}[1]{\protect\rule{.1in}{.1in}}

\begin{document}

\title{Spinorial coordinates for Lorentzian 4-metrics}
\author{D.C.Robinson\\Mathematics Department\\King's College London\\Strand\\London WC2R 2LS\\United Kingdom\\email: david.c.robinson@kcl.ac.uk}
\maketitle

\textbf{Abstract:}

Lorentzian 4-metrics are expressed in spinorial coordinates. \ In these
coordinates the metrics components can be factorized into a product of complex
conjugate quantities. \ The linearized theory and Einstein's vacuum field
equations are studied using these coordinates. \ The relationship between
Lorentzian and complex 4-metrics is discussed.\pagebreak

\section{Introduction}

In this paper a spinorial form for Lorentzian 4-metrics on real four
dimensional manifolds is presented and discussed. \ By using the two-component
spinor formalism \cite{pen1} it is shown that any Lorentzian 4-metric can be
locally expressed in terms of a conformal factor, a two index symmetric spinor
field and its complex conjugate. \ It is demonstrated that in this coordinate
system such a metric admits a factorization into a product of complex
conjugate quantities. \ These spinorial coordinates are used to study the
metric, and Einstein's vacuum field equations. \ The linear approximation
about flat space is also discussed.

The two-component spinor formalism that is needed is reviewed in the next two
sections. \ Here Cartan's structure equations for metric geometries and
Einstein's vacuum field equations for signature (1, 3) metrics are presented
and a linearized version of these equations is recalled. \ Anti-self dual (and
self-dual) solutions of Cartan's equations on real four dimensional manifolds
are defined. \ In the fourth and fifth sections spinor coordinates for
Lorentzian four-metrics are introduced and used to factorize the metric into a
product of complex conjugate quantities. \ These coordinates, and the
linearized versions of Cartan's structure equations and Einstein's vacuum
field equations, are used in the sixth section to derive the metric associated
with Roger Penrose's Hertz potentials for spin two fields in Minkowski
space-time \cite{pen2}. \ Similar coordinate systems arise in investigations
of holomorphic metrics, particularly in the approaches initiated by Jerzy
Pleba\'{n}ski and others, \cite{pleb1}, \cite{pleb3}. \ In the final section
spinorial coordinates and complex\ anti self-dual systems are considered and
their possible use in the construction of Ricci flat Lorentzian 4-metrics is
briefly discussed.

All considerations in this paper are local. \ Upper case Latin indices range
and sum over 0 to 1 and are raised and lowered with the antisymmetric spinors
$\epsilon^{AB}$ and $\epsilon_{AB}$ as in reference \cite{pen1}.

\section{Two component spinor formalism}

The spinor forms of Cartan's structure equations and Einstein's gravitational
equations used in this paper are as follows.

When the metric is written in terms of two component spinors so that%
\begin{equation}
ds^{2}=\epsilon_{AB}\epsilon_{A^{\prime}B^{\prime}}\theta^{AA^{\prime}}%
\otimes\theta^{BB^{\prime}},
\end{equation}
the first Cartan equations are%
\begin{equation}
D\theta^{AA^{\prime}}\equiv d\theta^{AA^{\prime}}+\omega_{B}^{A}%
\theta^{BA^{\prime}}+\overline{\omega}_{B^{\prime}}^{A^{\prime}}%
\theta^{AB^{\prime}}=0,
\end{equation}
where the co-frame $\theta^{AA^{\prime}}$ is a hermitian matrix-valued
one-form, the complex conjugate $\overline{\omega}_{B^{\prime}}^{A^{\prime}}$
and $\omega_{B}^{A}$ correspond, respectively, to the $sl(2,C)$-valued,
self-dual and anti self-dual parts of the torsion-free and metric connection
one form. \ The second Cartan equations are%
\begin{align}
\Omega_{B}^{A}  &  =d\omega_{B}^{A}+\omega_{C}^{A}\omega_{B}^{C}=\frac{1}%
{2}R_{BCC^{\prime}DD^{\prime}}^{A}\theta^{CC^{\prime}}\theta^{DD^{\prime}},\\
\overline{\Omega}_{B^{\prime}}^{A^{\prime}}  &  =d\overline{\omega}%
_{B^{\prime}}^{A^{\prime}}+\overline{\omega}_{C^{\prime}}^{A^{\prime}%
}\overline{\omega}_{B^{\prime}}^{C^{\prime}}=\frac{1}{2}\overline
{R}_{B^{\prime}CC^{\prime}DD^{\prime}}^{A^{\prime}}\theta^{CC^{\prime}}%
\theta^{DD^{\prime}},\nonumber
\end{align}
Here $\Omega_{B}^{A}$ and its complex conjugate $\overline{\Omega}_{B^{\prime
}}^{A^{\prime}}$ are $sl(2,C)$ valued two-forms and $R_{BCC^{\prime}%
DD^{\prime}}^{A}$ and its complex conjugate $\overline{R}_{B^{\prime
}CC^{\prime}DD^{\prime}}^{A^{\prime}}$ are, respectively, the anti-self dual
and self-dual parts of the Riemann tensor. Furthermore%
\begin{align}
\Omega_{B}^{A}  &  =\Psi_{BCD}^{A}\Sigma^{CD}+2\Lambda\Sigma_{B}^{A}%
+\Phi_{BC^{\prime}D^{\prime}}^{A}\text{ }\overline{\Sigma}^{C^{\prime
}D^{\prime}};\\
\overline{\Omega}_{B^{\prime}}^{A^{\prime}}  &  =\overline{\Psi}_{B^{\prime
}C^{\prime}D^{\prime}}^{A^{\prime}}\overline{\Sigma}^{C^{\prime}D^{\prime}%
}+2\Lambda\overline{\Sigma}_{B^{\prime}}^{A^{\prime}}+\Phi_{B^{\prime}%
CD}^{A^{\prime}}\overline{\Sigma}^{CD};\nonumber\\
\Sigma^{CD}  &  =\frac{1}{2}\theta_{^{B^{\prime}}}^{C}\theta^{DB^{\prime}%
}\text{, }\overline{\Sigma}^{C^{\prime}D^{\prime}}=\frac{1}{2}\theta
_{B}^{.^{C^{\prime}}}\theta^{BD^{\prime}}.\nonumber
\end{align}
The Weyl spinors $\Psi_{ABCD}=\Psi_{(ABCD)\text{ }}$ and $\overline{\Psi
}_{A^{\prime}B^{\prime}C^{\prime}D^{\prime}}=\overline{\Psi}_{(A^{\prime
}B^{\prime}C^{\prime}D^{\prime})\text{ }}$correspond respectively to the
anti-self dual and self-dual parts of the Weyl tensor, $-2\Phi_{BC^{\prime
}D^{\prime}}^{A}$ corresponds to the trace-free part of the Ricci tensor and
$24\Lambda$ corresponds to the Ricci scalar.

The first and second Bianchi identities are%
\begin{align}
\Omega_{B}^{A}\theta^{BA^{\prime}}+\overline{\Omega}_{B^{\prime}}^{A^{\prime}%
}\theta^{AB^{\prime}}  &  =0,\\
D\Omega_{B}^{A}  &  =D\overline{\Omega}_{B^{\prime}}^{A^{\prime}}=0,\nonumber
\end{align}
and $D$ always denotes the relevant covariant exterior derivative.

Under a change of co-frame%
\begin{align}
\theta^{AA^{\prime}}  &  \rightarrow(L^{-1})_{B}^{A}(L^{-1})_{B^{\prime}%
}^{A^{\prime}}\theta^{BB^{\prime}};\text{ }\omega_{B}^{A}\rightarrow
(L^{-1})_{C}^{A}dL_{B}^{C}+(L^{-1})_{C}^{A}\omega_{D}^{C}L_{B}^{D};\\
\Omega_{B}^{A}  &  \rightarrow(L^{-1})_{C}^{A}\Omega_{D}^{C}L_{B}^{D}\nonumber
\end{align}
where $L_{B}^{A}\in SL(2,C)$ and $\overline{L}_{B^{\prime}}^{A^{\prime}}$ is
its complex conjugate. \ Similar results hold for the complex conjugates
$\overline{\omega}_{B^{\prime}}^{A^{\prime}}$ and $\overline{\Omega
}_{B^{\prime}}^{A^{\prime}}$.

Einstein's vacuum field equations with zero cosmological constant,
$\Phi_{BC^{\prime}D^{\prime}}^{A}=\Lambda=0$, can also be written either as%
\begin{equation}
\Omega_{B}^{A}=\Psi_{BCD}^{A}\Sigma^{CD},
\end{equation}
or as%
\begin{equation}
\Omega_{B}^{A}\theta^{BA^{\prime}}=0.
\end{equation}
Cartan's equations can be extended to complex ones permitting complex
solutions and having as structure group $SO(4,C)$ $\sim SL(2,C)_{L}\times
SL(2,C)_{R}/\mathbb{Z}_{2}$. \ A (complex) anti-self dual solution of Cartan's
structure equations on a real four dimensional manifold is a complex co-frame
which satisfies the first Cartan structure equations above with flat self-dual
connection $\overline{\omega}_{B^{\prime}}^{A^{\prime}}$. \ The anti-self dual
curvature consequently satisfies Eqs.(7) and (8), that is $\Omega_{B}^{A}%
=\Psi_{BCD}^{A}\Sigma^{CD}$ and $\Omega_{B}^{A}\theta^{BA^{\prime}}=0.$ \ \ By
using Eq.(6) with $L_{B}^{C}\in SL(2,C)_{R}$ the self-dual connection forms
can be set equal to zero. \ Self-dual solutions are defined in an analogous
way. \ Anti-self dual solutions of Cartan's equations do not define real
four-metrics but they can be combined with their complex conjugates (that is
self-dual solutions) to construct Lorentzian 4-metrics. \ Examples of
combinations which are Ricci flat are discussed in the final section.

\section{Linearized equations}

Consider a metric linearized about the flat metric $ds^{2}=\epsilon
_{AB}\epsilon_{A^{\prime}B^{\prime}}dx^{AA^{\prime}}\otimes dx^{BB^{\prime}}$
so that the linearized metric takes the form%

\begin{equation}
ds_{lin}^{2}=(\epsilon_{AB}\epsilon_{A^{\prime}B^{\prime}}+\gamma_{AA^{\prime
}BB^{\prime}})dx^{AA^{\prime}}\otimes dx^{BB^{\prime}}.
\end{equation}
When a linearized co-frame is chosen to be%
\begin{equation}
\theta_{lin}^{AA^{\prime}}=(\delta_{B}^{A}\delta_{B^{\prime}}^{A^{\prime}}%
+\mu_{BB^{\prime}}^{AA^{\prime}})dx^{BB^{\prime}},
\end{equation}
so that%
\begin{align}
ds_{lin}^{2}  &  =\epsilon_{AB}\epsilon_{A^{\prime}B^{\prime}}\theta
_{lin}^{AA^{\prime}}\otimes\theta_{lin}^{BB^{\prime}}\\
&  =(\epsilon_{AB}\epsilon_{A^{\prime}B^{\prime}}+\mu_{AA^{\prime}BB^{\prime}%
}+\mu_{BB^{\prime}AA^{\prime}})dx^{AA^{\prime}}\otimes dx^{BB^{\prime}%
},\nonumber
\end{align}
then%
\begin{equation}
\gamma_{AA^{\prime}BB^{\prime}}=\mu_{AA^{\prime}BB^{\prime}}+\mu_{BB^{\prime
}AA^{\prime}}.
\end{equation}
The linearized Cartan equations for such a co-frame are \cite{mccul}%
\begin{align}
d\theta_{lin}^{AA^{\prime}}+\omega_{linB}^{A}dx^{BA^{\prime}}+\overline
{\omega}_{linB^{\prime}}^{A^{\prime}}dx^{AB^{\prime}}  &  =0;\text{ }\\
\Omega_{linB}^{A}  &  =d\omega_{linB}^{A},\nonumber
\end{align}
where%
\begin{align}
\Omega_{linB}^{A}  &  =\frac{1}{2}R_{linBCC^{\prime}DD^{\prime}}^{A}%
\epsilon^{CD}\overline{\Sigma}_{lin}^{C^{\prime}D^{\prime}}+\frac{1}%
{2}R_{linBCC^{\prime}DD^{\prime}}^{A}\epsilon^{C^{^{\prime}}D^{^{\prime}}%
}\Sigma_{lin}^{CD},\\
dx^{CC^{\prime}}dx^{DD^{\prime}}  &  =\epsilon^{CD}\overline{\Sigma}%
_{lin}^{C^{\prime}D^{\prime}}+\epsilon^{C^{^{\prime}}D^{^{\prime}}}%
\Sigma_{lin}^{CD},\nonumber
\end{align}
and similarly for the complex conjugates $\overline{\Omega}_{linB^{\prime}%
}^{A^{\prime}}$ and $d\overline{\omega}_{linB^{\prime}}^{A^{\prime}}$. \ The
linearized first and second Bianchi identities are%
\begin{align}
d\omega_{linB}^{A}dx^{BA^{\prime}}+d\overline{\omega}_{linB^{\prime}%
}^{A^{\prime}}dx^{AB^{\prime}}  &  =0,\\
d\Omega_{linB}^{A}  &  =\frac{1}{2}\partial_{EE^{\prime}}R_{linBCC^{\prime
}DD^{\prime}}^{A}dx^{CC^{\prime}}dx^{DD^{\prime}}dx^{EE^{\prime}%
}=0,\nonumber\\
d\overline{\Omega}_{linB^{\prime}}^{A^{\prime}}  &  =\frac{1}{2}%
\partial_{EE^{\prime}}\overline{R}_{linB^{\prime}CC^{\prime}DD^{\prime}%
}^{A^{\prime}}dx^{CC^{\prime}}dx^{DD^{\prime}}dx^{EE^{\prime}}=0.\nonumber
\end{align}

Under a linearized (first order) diffeomorphism%
\begin{equation}
x^{AA^{\prime}}\mapsto x^{AA^{\prime}}+\xi^{AA^{\prime}}%
\end{equation}%
\begin{align*}
\mu_{AA^{\prime}BB^{\prime}}  &  \mapsto\mu_{AA^{\prime}BB^{\prime}}%
+\partial_{BB^{\prime}}\xi_{AA^{\prime}},\\
\gamma_{AA^{\prime}BB^{\prime}}  &  \mapsto\gamma_{AA^{\prime}BB^{\prime}%
}+\partial_{BB^{\prime}}\xi_{AA^{\prime}}+\partial_{BB^{\prime}}%
\xi_{AA^{\prime}},
\end{align*}
where $\partial_{BB^{\prime}}$ denotes partial differentiation with respect to
$x^{BB^{\prime}}$. \ Under a linearized change of co-frame Eq.(6), linearized
about the identity, gives%
\begin{align}
\theta_{lin}^{AA^{\prime}}  &  \mapsto(\delta_{B}^{A}+l_{B}^{A})((\delta
_{B^{\prime}}^{A^{\prime}}+\overline{l}_{B^{\prime}}^{A^{\prime}})\theta
_{lin}^{BB^{\prime}},\\
l_{AB}  &  =l_{BA},\text{ }\overline{l}_{A^{\prime}B^{\prime}}=\overline
{l}_{B^{\prime}A^{\prime}},\nonumber
\end{align}
and%
\begin{equation}
\omega_{linB}^{A}\mapsto\omega_{linB}^{A}+dl_{B}^{A},\text{ }\overline{\omega
}_{linB^{\prime}}^{A^{\prime}}\mapsto\overline{\omega}_{linB^{\prime}%
}^{A^{\prime}}+d\overline{l}_{B^{\prime}}^{A^{\prime}}.
\end{equation}
It follows that a co-frame can be chosen for which $\mu_{AA^{\prime}%
BB^{\prime}}=\mu_{BB^{\prime}AA^{\prime}}$ and hence, with this choice,%
\begin{equation}
\gamma_{AA^{\prime}BB^{\prime}}=2\mu_{AA^{\prime}BB^{\prime}}.
\end{equation}
From Eqs.(7) and (8) the linearized Einstein vacuum field equations are%
\begin{equation}
d\omega_{linB}^{A}=\Omega_{linB}^{A}=\frac{1}{2}\Psi_{linBCD}^{A}%
dx_{D^{\prime}}^{C}dx^{DD^{\prime}},
\end{equation}
or%
\begin{equation}
\Omega_{linB}^{A}dx^{BB^{\prime}}=d(\omega_{linB}^{A}dx^{BB^{\prime}})=0,
\end{equation}
where $\Psi_{linBCD}^{A}$ denotes the components of the totally symmetric,
linearized anti-self dual Weyl spinor (and similarly for the linearized
self-dual quantities).

\section{Spinorial coordinates and Lorentzian 4-metrics}

Locally any Lorentzian 4-metric can always be written in terms of null
coordinates $(u,r,\zeta,\overline{\zeta})$%
\begin{equation}
ds^{2}=adu^{2}+2dudr+2bdud\zeta+2\overline{b}dud\overline{\zeta}+cd\zeta
^{2}+\overline{c}d\overline{\zeta}^{2}-2pd\zeta d\overline{\zeta.}%
\end{equation}
where here the hypersurfaces given by constant $u$ are chosen to be retarded
null hypersurfaces, $r$ is an affine parameter along the null geodesics ruling
such hypersurfaces and $\zeta$ is a complex (angular) coordinate labelling
such null geodesics. \ The latter can always be chosen, as will be done here,
so that $p>0$. \ Under a change of coordinates $r\rightarrow v$, where the
inverse transformation $r=r(u,v,\zeta,\overline{\zeta})$ is determined by the
equation%
\begin{equation}
\frac{\partial r}{\partial v}=p(u,r,\zeta,\overline{\zeta}),
\end{equation}
the metric takes the form%
\begin{equation}
ds^{2}=\exp2\sigma(2dudv-2d\zeta d\overline{\zeta}+Adu^{2}+2Bdud\zeta
+2\overline{B}dud\overline{\zeta}+Cd\zeta^{2}+\overline{C}d\overline{\zeta
}^{2})
\end{equation}
where%
\begin{align}
\exp2\sigma &  =p;\text{ }A=p^{-1}(a+2\frac{\partial r}{\partial u});\\
B  &  =p^{-1}(b+\frac{\partial r}{\partial\zeta});\text{ }C=cp^{-1}.\nonumber
\end{align}

The metric in Eq.(24) can be written in spinorial form by introducing
spinorial coordinates $x^{AA^{\prime}}$ and a spin dyad ($o^{A},\iota^{A})$
where
\begin{equation}
x^{AA^{\prime}}=\left[
\begin{array}
[c]{cc}%
v & \zeta\\
\overline{\zeta} & u
\end{array}
\right]  ;\text{ }o^{A}=\delta_{0}^{A},\text{ }\iota^{A}=\delta_{1.}^{A}%
\end{equation}
Then Eq.(24) takes the form%
\begin{align}
ds^{2}  &  =g_{AA^{\prime}BB^{\prime}}dx^{AA^{\prime}}\otimes dx^{BB^{\prime}%
}\\
&  =\exp(2\sigma)[\epsilon_{AB}\epsilon_{A^{\prime}B^{\prime}}+2o_{A}%
o_{B}\overline{\psi}_{A^{\prime}B^{\prime}}+2\overline{o}_{A^{\prime}%
}\overline{o}_{B^{\prime}}\psi_{AB}]dx^{AA^{\prime}}\otimes dx^{BB^{\prime}%
},\nonumber
\end{align}
where the symmetric spinor $\psi_{AB}$ is related to $A$, $B$ and $C$ by%
\begin{align}
\psi_{00}  &  =\frac{C}{2},\text{ }\psi_{01}=\frac{B}{2},\text{ }\psi
_{11}+\overline{\psi}_{11}=\frac{A}{2},\nonumber\\
\psi_{AB}  &  =\psi_{00}\iota_{A}\iota_{B}-\psi_{01}(o_{A}\iota_{B}+\iota
_{A}o_{B})+\psi_{11}o_{A}o_{B}.
\end{align}

This metric form is preserved by the global coordinate transformations
\begin{align}
x^{AA^{\prime}}  &  \mapsto e^{-\lambda}L_{B}^{A}\overline{L}_{B^{\prime}%
}^{A^{\prime}}x^{BB^{\prime}}+p^{AA^{\prime}},\\
L_{C}^{A}L_{D}^{B}\epsilon_{AB}  &  =\epsilon_{CD},\nonumber
\end{align}
where \textit{here} $\lambda,L_{B}^{A}$ and $p^{AA^{\prime}}$ are constants,
and%
\begin{align}
\sigma &  \mapsto\sigma+\lambda,o_{A}\mapsto(L^{-1})_{A}^{B}o_{B},\\
\psi_{AB}  &  \mapsto(L^{-1})_{A}^{C}(L^{-1})_{B}^{D}\psi_{CD},\nonumber
\end{align}
and similarly for the complex conjugates. \ Furthermore $o^{A}\mapsto
\delta_{0}^{A}$ when $L_{0}^{A}=\delta_{0}^{A}$. \ In addition this form of
the metric is preserved under the transformation $x^{AA^{\prime}}\mapsto
x^{AA^{\prime}}+2o^{A}\overline{o}^{A^{\prime}}f(u)$, $\psi_{AB}\mapsto
\psi_{AB}+o_{A}o_{B}\frac{df}{du}$, for any real-valued function $f$.

The conformal geometry is determined by the symmetric spinor $\psi_{AB}$ but
it should be noted that $\operatorname{Im}\psi_{11}$ does not appear in the
conformal metric.

The inverse $g^{-1}$of the metric $g$ with components $g_{AA^{\prime
}BB^{\prime}}$ has components%
\begin{align}
(g^{^{-1}})^{AA^{\prime}BB^{\prime}}  &  =\exp(-2\sigma)[1-4\psi_{00}%
\overline{\psi}_{0^{\prime}0^{\prime}}]^{-1}[\epsilon^{AB}\epsilon^{A^{\prime
}B^{\prime}}-2o^{A}o^{B}\overline{\psi}^{A^{\prime}B^{\prime}}-2\overline
{o}^{A^{\prime}}\overline{o}^{B^{\prime}}\psi^{AB}\\
&  -4o^{A}\overline{o}^{A^{\prime}}\psi^{B}\overline{\psi}^{B^{\prime}}%
-4o^{B}\overline{o}^{B^{\prime}}\psi^{A}\overline{\psi}^{A^{\prime}}%
+4o^{A}\overline{o}^{A^{\prime}}o^{B}\overline{o}^{B^{\prime}}(\Delta
\overline{\psi}_{0^{\prime}0^{\prime}}+\overline{\Delta}\psi_{00})]\nonumber
\end{align}

where%
\begin{align}
\psi^{A}  &  =\psi_{B}^{A}o^{B}=\psi_{0}^{A}\\
\Delta &  =\psi_{AB}\psi^{AB},\nonumber
\end{align}
and similarly for the complex conjugate quantities. Consequently regularity of
the inverse requires $4\psi_{00}\overline{\psi}_{0^{\prime}0^{\prime}}\neq1$.

A co-frame for the metric given in Eq.(27) is%
\begin{equation}
\theta^{AA^{\prime}}=\exp(\sigma-\zeta)[\delta_{B}^{A}\delta_{B^{\prime}%
}^{A^{\prime}}+o^{A}o_{B}\overline{\varphi}_{B^{\prime}}^{A^{\prime}%
}+\overline{o}^{A^{\prime}}\overline{o}_{B^{\prime}}\varphi_{B}^{A}%
+o^{A}\overline{o}^{A^{\prime}}\varphi_{0B}\overline{\varphi}_{0^{\prime
}B^{\prime}}]dx^{BB^{\prime}}%
\end{equation}

where%
\begin{align}
\varphi_{AB}  &  =\psi_{AB}\exp2\zeta,\\
\exp2\zeta &  =1+\varphi_{00}\overline{\varphi}_{0^{\prime}0^{\prime}%
}.\nonumber
\end{align}
and
\[
ds^{2}=\epsilon_{AB}\epsilon_{A^{\prime}B^{\prime}}\theta^{AA^{\prime}}%
\otimes\theta^{BB^{\prime}}.
\]
The dual frame is%
\begin{align}
E_{BB^{\prime}}  &  =[\exp(\zeta-\sigma)][(1-\varphi\overline{\varphi}%
)^{-1}][\delta_{B}^{A}\delta_{B^{\prime}}^{A^{\prime}}-o^{A}o_{B}%
\overline{\varphi}_{B^{\prime}}^{A^{\prime}}-\overline{o}^{A^{\prime}%
}\overline{o}_{B^{\prime}}\varphi_{B}^{A}\\
&  -o_{B}\overline{o}_{B^{\prime}}\varphi_{0}^{A}\overline{\varphi}%
_{0^{\prime}}^{A^{\prime}}-2(1+\varphi\overline{\varphi})^{-1}o^{A}%
\overline{o}^{A^{\prime}}\varphi_{B0}\overline{\varphi}_{B^{\prime}0^{\prime}%
}\nonumber\\
&  +(1+\varphi\overline{\varphi})^{-1}(\varphi\overline{\Delta}_{\varphi
}+\overline{\varphi}\Delta_{\varphi})o^{A}\overline{o}^{A^{\prime}}%
o_{B}\overline{o}_{B^{\prime}}]\frac{\partial}{\partial x^{AA^{\prime}}%
},\nonumber
\end{align}

where $\varphi=\varphi_{00},\Delta_{\varphi}=\varphi_{AB}\varphi^{AB}$ and
similarly for the complex conjugate quantities.

\section{Factorization of co-frame and metric}

An interesting feature of the metric in spinor coordinates as presented in the
previous section is that the expressions for the co-frame and metric given in
Eqs.(33)\ and (27) admit factorizations into the products of complex conjugate
terms as follows. \ The co-frame factorizes as%
\begin{equation}
\theta^{AA^{\prime}}=\chi_{BP^{\prime}}^{PA^{\prime}}\overline{\chi
}_{PB^{\prime}}^{AP^{\prime}}dx^{BB^{\prime}}.
\end{equation}
where%
\begin{equation}
\chi_{BP^{\prime}}^{PA^{\prime}}=\exp\alpha(\delta_{B}^{P}\delta_{P^{\prime}%
}^{A^{\prime}}+\overline{o}^{A^{\prime}}\overline{o}_{P^{\prime}}\varphi
_{B}^{P}),
\end{equation}
and%
\begin{equation}
\alpha+\overline{\alpha}=\exp(\sigma-\zeta).
\end{equation}
It follows that the metric components given in Eq.(27) can also be written as
the product of complex conjugate terms. \ If this metric's components are
written as $g_{CC^{\prime}DD^{\prime}}$ so that Eq.(27) is%
\begin{equation}
ds^{2}=g_{CC^{\prime}DD^{\prime}}dx^{CC^{\prime}}\otimes dx^{DD^{\prime}},
\end{equation}
then it follows that%
\begin{equation}
g_{CC^{\prime}DD^{\prime}}=k^{P}{}_{CP^{\prime}}._{.DQ^{\prime}}^{Q}%
\overline{k}^{P^{\prime}}{}_{C^{\prime}P}._{.D^{\prime}Q}^{Q^{\prime}}%
\end{equation}
where%
\begin{equation}
k_{PCP^{\prime}QDQ^{\prime}}=\epsilon_{A^{\prime}B^{\prime}}\chi_{PCP^{\prime
}}^{A^{\prime}}\chi_{QDQ^{\prime}}^{B^{\prime}},
\end{equation}
so%
\begin{equation}
k_{PCP^{\prime}QDQ^{\prime}}=\frac{1}{2}\exp2\alpha\lbrack\epsilon
_{PC}h_{QP^{\prime}DQ^{\prime}}+\epsilon_{DQ}h_{CP^{\prime}PQ^{\prime}}],
\end{equation}
where
\begin{equation}
h_{QP^{\prime}DQ^{\prime}}=\epsilon_{QD}\epsilon_{P^{\prime}Q^{\prime}%
}+2\overline{o}_{P^{\prime}}\overline{o}_{Q^{\prime}}\varphi_{QP,}%
\end{equation}
and similarly for the complex conjugate of $k_{PCP^{\prime}QDQ^{\prime}}$.
\ The imaginary part of $\alpha$ does not appear in the metric, Eq.(27), which
has components $g_{CC^{\prime}DD^{\prime}}$ equal to%
\begin{equation}
\frac{1}{4}\exp(2\alpha+2\overline{\alpha})[h_{C^{\prime}DQ^{\prime}}%
^{Q}\overline{h}_{CD^{\prime}Q}^{Q^{\prime}}-h_{Q^{\prime}DD^{\prime}}%
^{Q}\overline{h}_{CC^{\prime}Q}{}^{Q^{\prime}}-h_{CC^{\prime}}{}_{Q^{\prime}%
}^{Q}\overline{h}_{QDD^{\prime}}^{Q^{\prime}}+h_{CQ^{\prime}}{}_{D^{\prime}%
}^{Q}\overline{h}_{QC^{\prime}D}\text{ }^{Q^{\prime}}].
\end{equation}

\section{Linearization of the 4-metric using spinorial coordinates}

Now consider again the metric in Eq.(27) but now linearized about the
Minkowski metric so that%
\begin{align}
ds_{lin}^{2}  &  =(\epsilon_{AB}\epsilon_{A^{\prime}B^{\prime}}+\gamma
_{AA^{\prime}BB^{\prime}})dx^{AA^{\prime}}\otimes dx^{BB^{\prime}},\\
\gamma_{AA^{\prime}BB^{\prime}}  &  =2\sigma\epsilon_{AB}\epsilon_{A^{\prime
}B^{\prime}}+2o_{A}o_{B}\overline{\psi}_{A^{\prime}B^{\prime}}+2\overline
{o}_{A^{\prime}}\overline{o}_{B^{\prime}}\psi_{AB},\nonumber
\end{align}
where \textit{in this section only} $\sigma$ and $\psi_{AB}=\varphi_{AB}$ are
first order terms. \ Hence a linearized co-frame and frame for the linearized
metric are\footnote{Only a restricted set of the linearized diffeomorphisms of
Eq.(16) satisfying $\partial_{BB^{\prime}}\xi_{AA^{\prime}}=\partial
_{BB^{\prime}}\partial_{AA^{\prime}}\xi$ $=\frac{1}{4}\square\xi\epsilon
_{AB}\epsilon_{A^{\prime}B^{\prime}}+\overline{o}_{A^{\prime}}\overline
{o}_{B^{\prime}}\zeta_{AB}+o_{A}o_{B}\overline{\zeta}_{A^{\prime}B^{\prime}}$,
for a real function $\xi$, and the complex conjugates $\zeta_{AB}=\zeta_{BA}$
and $\overline{\zeta}_{A^{\prime}B^{\prime}}=\overline{\zeta}_{B^{\prime
}A^{\prime}},$ preserve the form of both this metric and these bases.}%
\begin{align}
\theta_{lin}^{AA^{\prime}}  &  =[\delta_{B}^{A}\delta_{B^{\prime}}^{A^{\prime
}}(1+\sigma)+o^{A}o_{B}\overline{\psi}_{B^{\prime}}^{A^{\prime}}+\overline
{o}^{A^{\prime}}\overline{o}_{B^{\prime}}\psi_{B}^{A}]dx^{BB^{\prime}},\\
E_{linBB^{\prime}}  &  =(\delta_{B}^{A}\delta_{B^{\prime}}^{A^{\prime}%
}(1-\sigma)-o^{A}o_{B}\overline{\psi}_{B^{\prime}}^{A^{\prime}}-\overline
{o}^{A^{\prime}}\overline{o}_{B^{\prime}}\psi_{B}^{A})\frac{\partial}{\partial
x^{AA^{\prime}}}.\nonumber
\end{align}

It follows from Eq.(13) that the corresponding anti-self dual linearized
connection one-form is given by%
\begin{align}
\omega_{linAB}  &  =\frac{1}{2}[-\partial_{AC^{\prime}}\sigma\epsilon
_{BC}-\partial_{BC^{\prime}}\sigma\epsilon_{AC}+o_{C}(o_{A}\partial
_{BA^{\prime}}\overline{\psi}_{C^{\prime}}^{A^{\prime}}+o_{B}\partial
_{AA^{\prime}}\overline{\psi}_{C^{\prime}}^{A^{\prime}})\\
&  +\overline{o}_{C^{\prime}}\overline{o}^{A^{\prime}}(\partial_{BA^{\prime}%
}\psi_{AC}+\partial_{AA^{\prime}}\psi_{BC})]dx^{CC^{\prime}}\nonumber
\end{align}
and similarly for the complex conjugate self-dual connection one-form. \ The
anti-self dual curvature two-forms is given by%
\begin{align}
&  R_{linABCC^{\prime}DD^{\prime}}dx^{CC^{\prime}}dx^{DD^{\prime}}\\
&  =[-\partial_{(B\mid C^{\prime}\mid}\partial_{A)D^{\prime}}\sigma
-\partial_{(B\mid D^{\prime}\mid}\partial_{A)C^{\prime}}\sigma\nonumber\\
&  +o^{C}(o_{(A}\partial_{B)A^{\prime}}\partial_{CD^{\prime}}\overline{\psi
}_{C^{\prime}}^{A^{\prime}}+o_{(A}\partial_{B)A^{\prime}}\partial_{CC^{\prime
}}\overline{\psi}_{D^{\prime}}^{A^{\prime}})\nonumber\\
&  -\overline{o}^{A^{\prime}}(\overline{o}_{C^{\prime}}\partial_{D^{\prime}%
}^{C}\partial_{A^{\prime}(B}\psi_{A)C}+\overline{o}_{D^{\prime}}%
\partial_{C^{\prime}}^{C}\partial_{A^{\prime}(B}\psi_{A)C})]\overline{\Sigma
}_{lin}^{C^{\prime}D^{\prime}}\nonumber\\
&  +[\frac{1}{2}(\epsilon_{AC}\epsilon_{BD}+\epsilon_{BC}\epsilon_{AD}%
)\square\sigma+o_{A}o_{(C}\partial_{D)C^{\prime}}\partial_{BA^{\prime}%
}\overline{\psi}^{A^{\prime}C^{\prime}}\nonumber\\
&  +o_{B}o_{(C}\partial_{D)C^{\prime}}\partial_{AA^{\prime}}\overline{\psi
}^{A^{\prime}C^{\prime}}+\overline{o}^{A^{\prime}}\overline{o}^{C^{\prime}%
}\partial_{DC^{\prime}}\partial_{A^{\prime}(B}\psi_{A)C}+\overline
{o}^{A^{\prime}}\overline{o}^{C^{\prime}}\partial_{CC^{\prime}}\partial
_{A^{\prime}(B}\psi_{A)D}]\Sigma_{lin}^{CD}.\nonumber
\end{align}
where $\partial_{AA^{\prime}}\partial_{B}^{A^{\prime}}=\frac{1}{2}%
\epsilon_{AB}\square$, and similarly for the complex conjugate self-dual
curvature. \ While this expression is more complicated than the expression for
the linearized curvature obtained in the standard way a number of interesting
conclusions can be drawn from it. \ For instance, comparing it with the
equations of Sec.3 it can immediately be seen that the linearized curvature
has vanishing Ricci tensor when $\sigma=0$ and $\psi_{AB}$ satisfies the
source-free Maxwell equations in Minkowski space-time, that is%
\begin{equation}
\partial_{AA^{\prime}}\psi^{AC}=0.
\end{equation}
In this case the anti-self dual linearized Weyl spinor is given by
$\Psi_{linABCD}=\frac{1}{2}[\partial_{D}\partial_{(B}\psi_{A)C}+\partial
_{C}\partial_{(B}\psi_{A)D}],$ where $\overline{o}^{A^{\prime}}\partial
_{AA^{\prime}}\equiv\partial_{A}$, and similarly for the complex conjugate
self-dual Weyl spinor.

Many years ago Penrose, \cite{pen1}, showed locally, and globally subject to
topological conditions, that a totally symmetric spinor, $\Xi_{ABCD..P}$,
satisfies the spin $s$ zero rest-mass field equation in Minkowski space-time%
\begin{equation}
\partial_{AA^{\prime}}\Xi_{BCD..P}^{A}=0
\end{equation}
if and only if, for any choice of a constant spinor $\kappa^{A^{\prime}}$,
there exists a complex function $\xi$ such that%
\begin{equation}
\Xi_{ABCD..P}=\kappa^{A^{\prime}}\kappa^{B^{\prime}}\kappa^{C^{\prime}}%
\kappa^{D^{\prime}}..\kappa^{P^{\prime}}\partial_{AA^{\prime}}\partial
_{BB^{\prime}}\partial_{CC^{\prime}}\partial_{DD^{\prime}...}\partial
_{PP^{\prime}}\xi
\end{equation}
where $\xi$ satisfies the wave equation,%
\begin{equation}
\square\xi=0.
\end{equation}

When $s=2$, with $\Xi_{ABCD}=\Psi_{linABCD}$, Eq.(50) corresponds to the
linearized second Bianchi identity when the linearized Einstein vacuum field
equations are satisfied. \ In this case, with $\kappa^{A^{\prime}}%
=\overline{o}^{A^{\prime}},$ Eqs.(51) and (52) become%
\begin{align}
\Psi_{linABCD}  &  =\partial_{A}\partial_{B}\partial_{C}\partial_{D}\psi
_{lin},\\
\square\psi_{lin}  &  =0.\nonumber
\end{align}
By using these results and the expression $\Psi_{linABCD}=\partial_{A}%
\partial_{B}\partial_{C}\partial_{D}\psi_{lin}$ in the second linearized
Cartan equation of Eq.(13), the corresponding linearized anti self-dual
connection one-form can be shown to be%
\begin{equation}
\omega_{linAB}=\partial_{A}\partial_{B}\partial_{C}\psi dx^{CC^{\prime}%
}\overline{o}_{C^{\prime}}+d\alpha_{AB}%
\end{equation}
Here $\alpha_{AB}$ are arbitrary functions which can be removed by using the
linearized gauge transformation Eq.(18). \ \ It then follows, by using this
result and its complex conjugate in the first linearized Cartan equation,
Eq.(13), that%
\begin{align}
\theta_{lin}^{AA^{\prime}}  &  =(\delta_{B}^{A}\delta_{B^{\prime}}^{A^{\prime
}}+o^{A}o_{B}\overline{\psi}_{B^{\prime}}^{A^{\prime}}+\overline{o}%
^{A^{\prime}}\overline{o}_{B^{\prime}}\psi_{B}^{A}+\partial_{BB^{\prime}}%
\beta^{AA^{\prime}})dx^{BB^{\prime}},\\
\psi_{AB}  &  =\partial_{A}\partial_{B}\psi;\text{ }\overline{\psi}%
_{A^{\prime}B^{\prime}}=\partial_{A^{\prime}}\partial_{B^{\prime}}%
\overline{\psi},\nonumber
\end{align}
where $\beta^{AA^{\prime}}$ are arbitrary functions which can be set equal to
zero by using the linearized diffeomorphisms of Eq.(16).

Hence Penrose's result implies that the linearized Einstein vacuum field
equations are satisfied by the linearized metric in Eq.(45) when $\sigma=0$
and the complex function $\psi$ satisfies the Minkowski space-time wave
equation. \ The linearized vacuum solutions are then given by%
\begin{align}
ds_{lin}^{2}  &  =(\epsilon_{AB}\epsilon_{A^{\prime}B^{\prime}}+2o_{A}%
o_{B}\partial_{A^{\prime}}\partial_{B^{\prime}}\overline{\psi}+2\overline
{o}_{A^{\prime}}\overline{o}_{B^{\prime}}\partial_{A}\partial_{B}%
\psi)dx^{AA^{\prime}}\otimes dx^{BB^{\prime}}],\\
\square\psi &  =\partial^{AA^{\prime}}\partial_{AA^{\prime}}\psi=0.\nonumber
\end{align}
Similar linearized solutions were identified by Jerzy Pleba\'{n}ski and Ivor
Robinson working in the complex domain \cite{pleb2}. \ A useful discussion of
other work on Hertz potentials is included in \cite{stew}.

Finally in this section it should be noted that, in the factorization,
considered in the previous section, $\alpha+\overline{\alpha}=1+\sigma$ and
$\varphi=\psi,$ when only zeroth and first order terms are retained in these expressions.

\section{Complex and real solutions}

Holomorphic 4-metrics on complex four dimensional manifolds have been
extensively investigated, particularly in the context of half-flat metrics as
in the approach of Newman,\cite{newman}, the use of twistors,\cite{pen3} and
the work of Pleba\'{n}ski,\cite{pleb1}. \ A selection of reviews of this
research can be found in \cite{Dunajski},\cite{Ko},\cite{mason}. \ The aim of
this section is to discuss the relationship between complex and real solutions
of Cartan's and Einstein's equations using work on holomorphic metrics,
spinorial coordinates, and the results of Sec. 5.

As far as this paper is concerned certain complex solutions of Cartan's
structure equations on a real four dimensional manifold can be simply obtained
by re-interpreting formulae obtained by Pleba\'{n}ski and co-workers in their
research on holomorphic half-flat metrics\cite{pleb1, pleb2, pleb3}%
\footnote{Subsequently it was realized that research on Wave Geometry in
Hiroshima in the 1930's predates some of this work. \ It is reviewed in
\cite{mimura}.}. \ In particular certain complex anti self-dual (or self-dual)
solutions to Eqs.(2-4) and Eqs.(7-8) can be so obtained. \ These solutions are
given, in spinorial coordinates $x^{AA^{\prime}}$, by the complex one-forms
$\chi^{AA^{\prime}}=\chi_{BB^{\prime}}^{AA^{\prime}}dx^{BB^{\prime}}$,
constructed using Eq.(37) but with $\alpha=0$, $\overline{\varphi}_{A^{\prime
}B^{\prime}}=0$ and $\varphi_{AB}=\psi_{AB}=\partial_{A}\partial_{B}\psi$.
\ These complex one-forms satisfy the first set of Cartan's equations, Eq.(2),
when $\psi$ satisfies the generalized wave equation%
\begin{equation}
\square\psi-\psi_{AB}\psi^{AB}=0.
\end{equation}
\ Eqs. (7) and (8) are then automatically satisfied. \ The complex co-frame,
connection one-forms and the anti-self dual curvature two-forms satisfying
Cartan's equations are then%
\begin{align}
\theta^{AA^{\prime}}  &  =\chi^{AA^{\prime}},\\
\overline{\omega}_{A^{\prime}B^{\prime}}  &  =0,\text{ }\omega_{AB}%
=\overline{o}_{C^{\prime}}\partial_{A}\partial_{B}\partial_{C}\psi
dx^{CC^{\prime}},\nonumber\\
\Omega_{B}^{A}  &  =\partial^{A}\partial_{B}\partial_{C}\partial_{D}\psi
\Sigma^{CD}.\nonumber
\end{align}
The corresponding spinorial quantities $h_{QP^{\prime}DQ^{\prime}}$ and
$\overline{h}_{QP^{\prime}DQ^{\prime}}$ of Sec. (5) are no longer complex
conjugates and are given by
\begin{align}
h_{QP^{\prime}DQ^{\prime}}  &  =\epsilon_{QD}\epsilon_{P^{\prime}Q^{\prime}%
}+2\overline{o}_{P^{\prime}}\overline{o}_{Q^{\prime}}\partial_{A}\partial
_{B}\psi,\\
\overline{h}_{QP^{\prime}DQ^{\prime}}  &  =\epsilon_{QD}\epsilon_{P^{\prime
}Q^{\prime}}.\nonumber
\end{align}

The relationship, mentioned above, of these equations to those formulated in
the holomorphic context is the following. \ Introducing certain complex
coordinates, denoted here as complex spinorial coordinates $z^{AA^{\prime}}$,
Pleba\'{n}ski showed that all half-flat holomorphic four-metrics on complex
four dimensional manifolds could be locally expressed as%
\begin{equation}
ds^{2}=h_{AA^{\prime}BB^{\prime}\text{ }}dz^{AA^{\prime}}\otimes
dz^{BB^{\prime}},
\end{equation}
where the holomorphic metric components $h_{AA^{\prime}BB^{\prime}\text{ }}%
$are given by the holomorphic version of Eq.(43) with $\varphi_{AB}=\psi
_{AB}=\overline{o}^{A^{\prime}}\overline{o}^{B^{\prime}}\partial
/z_{AA^{\prime}}\partial/z_{BB^{\prime}}\psi$. \ \ \ The holomorphic version
of Eq.(57) is Pleba\'{n}ski's second heavenly equation \cite{pleb1}. \ If
$z^{AA^{\prime}}$ $=x^{AA^{\prime}}+iy^{AA^{\prime}}$, with $x^{AA^{\prime}}%
$and $\ y^{AA^{\prime}}$ the spinor correspondents of real coordinates $x^{a}$
and $y^{a}$, the pullbacks of Pleba\'{n}ski's holomorphic forms to the real
four manifold $M$ given by $y^{a}=0$ gives a class of anti self-dual solutions
of the complex Cartan equations on $M$ (and similarly for self-dual solutions).

It has been demonstrated in other papers that, by using spinorial coordinates
and the results of Sec.5, certain real solutions of Einstein's vacuum field
equations can be constructed \cite{rob2}, \cite{rob3}. \ These satisfy the
calculationally simplifying condition that $o^{A}$ is a principal spinor of
$\psi_{AB}$. \ When anti self-dual solutions, satisfying both this condtion
and Eq.(57) are combined with their complex conjugates, using the co-frame
$\chi^{AA^{\prime}}$ and its complex conjugate as in\ Eqs.(36) and (37) in
Sec.5, Ricci flat Lorentzian metrics result. \ \ A co-frame for these real
metrics is given by Eq.(33), or equivalently the combination of self-dual and
anti self-dual expressions in Eqs.(36) and (37), with $\varphi_{AB}=\psi_{AB}$
and $\sigma=0$. \ The Lorentzian line elements of these solutions are given by
Eq.(27), with $\sigma=0$ and $\psi_{AB}=\partial_{A}\partial_{B}\psi$ (plus
its complex conjugate). \ These vacuum solutions are Petrov type III or N.

Can further interesting Lorentzian metrics be constructed by combining complex
solutions, either by using spinorial coordinates and the approach of Sec.5 or
in some other way? \ To date this question has received only limited and
partial answers. \ A discussion of some other answers can be found in
\cite{poles}.

In conclusion it should be noted that spinorial coordinates may have other
uses. \ For instance they may be a useful tool in the analysis of
asymptotically flat metrics and radiating systems.\bigskip

\textbf{Acknowledgement}: \ I thank Maciej Dunajski for some references.
\ Helpful discussions with the late Ed Glass are gratefully
remembered.\pagebreak

\end{document}